\documentclass[footinbib,a4paper,aps,prl,reprint,twocolumn,preprintnumbers,amsmath,amssymb,10pt, superscriptaddress]{revtex4-1}
\usepackage{amsmath}
\usepackage{verbatim}
\usepackage{graphicx}
\usepackage{color}
\usepackage{tikz}
\usepackage[colorlinks=true,citecolor=blue,linkcolor=blue,urlcolor=blue]{hyperref}
\usepackage{braket, enumitem,}
\usepackage[normalem]{ulem}

\begin{document}

\title{Confinement and lack of thermalization after quenches in the bosonic Schwinger model}

\author{Titas Chanda}
\email{titas.chanda@uj.edu.pl}
\affiliation{Instytut Fizyki im. Mariana Smoluchowskiego, Uniwersytet Jagiello\'nski, \L{}ojasiewicza 11, 30-348 Krak\'ow, Poland} 

\author{Jakub Zakrzewski}
\affiliation{Instytut Fizyki im. Mariana Smoluchowskiego, Uniwersytet Jagiello\'nski, \L{}ojasiewicza 11, 30-348 Krak\'ow, Poland} 
\affiliation{Mark Kac Complex Systems Research Center,  Jagiellonian University in Krakow,  \L{}ojasiewicza 11, 30-348 Krak{\'o}w, Poland}

\author{Maciej Lewenstein}
\affiliation{ICFO-Institut de Ci\`encies Fot\`oniques, The Barcelona Institute of Science and Technology, Av. Carl Friedrich
Gauss 3, 08860 Castelldefels (Barcelona), Spain}
\affiliation{ICREA, Passeig Lluis Companys 23, 08010 Barcelona, Spain}

\author{Luca Tagliacozzo}
\affiliation{Department of Physics and SUPA, University of Strathclyde, Glasgow G4 0NG, UK}
\affiliation{Department de F\'{\i}sica Qu\`antica i Astrof\'{\i}sica and Institut de Ci\`encies del Cosmos (ICCUB), Universitat de Barcelona,  Mart\'{\i} i Franqu\`es 1, 08028 Barcelona, Catalonia, Spain}

\begin{abstract}
We excite the vacuum  of a relativistic theory of bosons coupled to a $U(1)$ gauge field in $1+1$ dimensions (bosonic Schwinger model) out of equilibrium  by creating a spatially separated particle-antiparticle pair connected by a string of electric field. During the evolution, we observe a strong confinement of  bosons witnessed by the bending of their light cone, reminiscent of what has been observed for the Ising model  [Nat. Phys. {\bf 13}, 246 (2017)]. As a consequence, for the time scales we are able to simulate, the system evades thermalization  and generates exotic asymptotic states. These states are made of two disjoint regions, an external deconfined region that seems to thermalize, and an inner core that  
reveals an area-law saturation of the entanglement entropy.
\end{abstract}
\maketitle

\paragraph{Introduction.--}
Solving the  out-of-equilibrium dynamics (OED) of large many-body quantum systems becomes  exponentially hard when the number of constituents increases beyond few tens. This fact hinders our understanding of  important questions such as the existence of new phases of matter \cite{fausti_science_2011, abanin_adp_2017, alet_crp_2018} and the presence or absence of thermalization \cite{rigol_nature_2008, polkovnikov_rmp_2011, eisert_natphys_2015, dalessio_ap_2016}.

Symmetries play a crucial role out of equilibrium as they give rise to conservation laws and continuity equations that can strongly constrain the dynamics  \cite{noether1918, castro-alvaredo_prx_2016, bertini_prl_2016}. Local symmetries can also have strong consequences in the OED since they are responsible for interesting phenomena, such as slow dynamics and localization \cite{brenes_prl_2018, robinson_prb_2019, james_prl_2019, surace_arxiv_2019, park_pra_2019, cubero_arxiv_2019}. 

The simplest system with local symmetries is the fermionic Schwinger model (FSM) \cite{schwinger_pr_1951}, the quantum electrodynamics in 1+1 dimensions (1D). The FSM, by now, has been studied extensively with traditional methods \cite{schwinger_pr_1962, schwinger_pr_1962_2, coleman_aop_1976, hamer_npb_1982, adam_aop_1997,brenes_prl_2018, zache_prl_2019} and with tensor-network techniques \cite{byrnes_prd_2002, banuls_jhep_2013, buyens_prl_2014, kuhn_pra_2014, banuls_prd_2015, buyens_prd_2016, pichler_prx_2016, banuls_prl_2017, ercolessi_prd_2018, huang_prl_2019, magnifico_arxiv_2019}. Despite its simplicity, the FSM shares common features with quantum chromodynamics (QCD), such as a non-trivial vacuum exhibiting chiral symmetry breaking and confinement. In particular, the FSM  confinement can be seen emerging from a quadratic perturbation of the sine-Gordon model, producing a similar effect to a magnetic field  in the Ising model \cite{delfino_NPB_1996, delfino_NPB_1998, mussardo_JHEP_2007}.   

Here, we consider the OED of the bosonic version of the Schwinger model \cite{schwinger_pr_1951, schwinger_pr_1962, schwinger_pr_1962_2} (BSM),
since the bosonic versions of well studied fermionic models can unveil unexpected new phenomena \cite{gonzalez_cuadra_prl_2018, gonzalez_cuadra_prd_2019, gonzalez_cuadra_natcomm_2019}. Moreover, the BSM is easier to realize in experiments with ultra-cold atoms \cite{kasamatsu_prl_2013, dutta_pra_2017, gonzalez_cuadra_njp_2017, kuno_prd_2017} using the tools already available in the labs \cite{nascimbene_prl_2012, martinez_nature_2016, bernien_nature_2017, kico_pra_2018, gorg_natphys_2019, schweizer_arxiv_2019, mil_arxiv_2019}.

\begin{figure}[t]
\includegraphics[width=\linewidth]{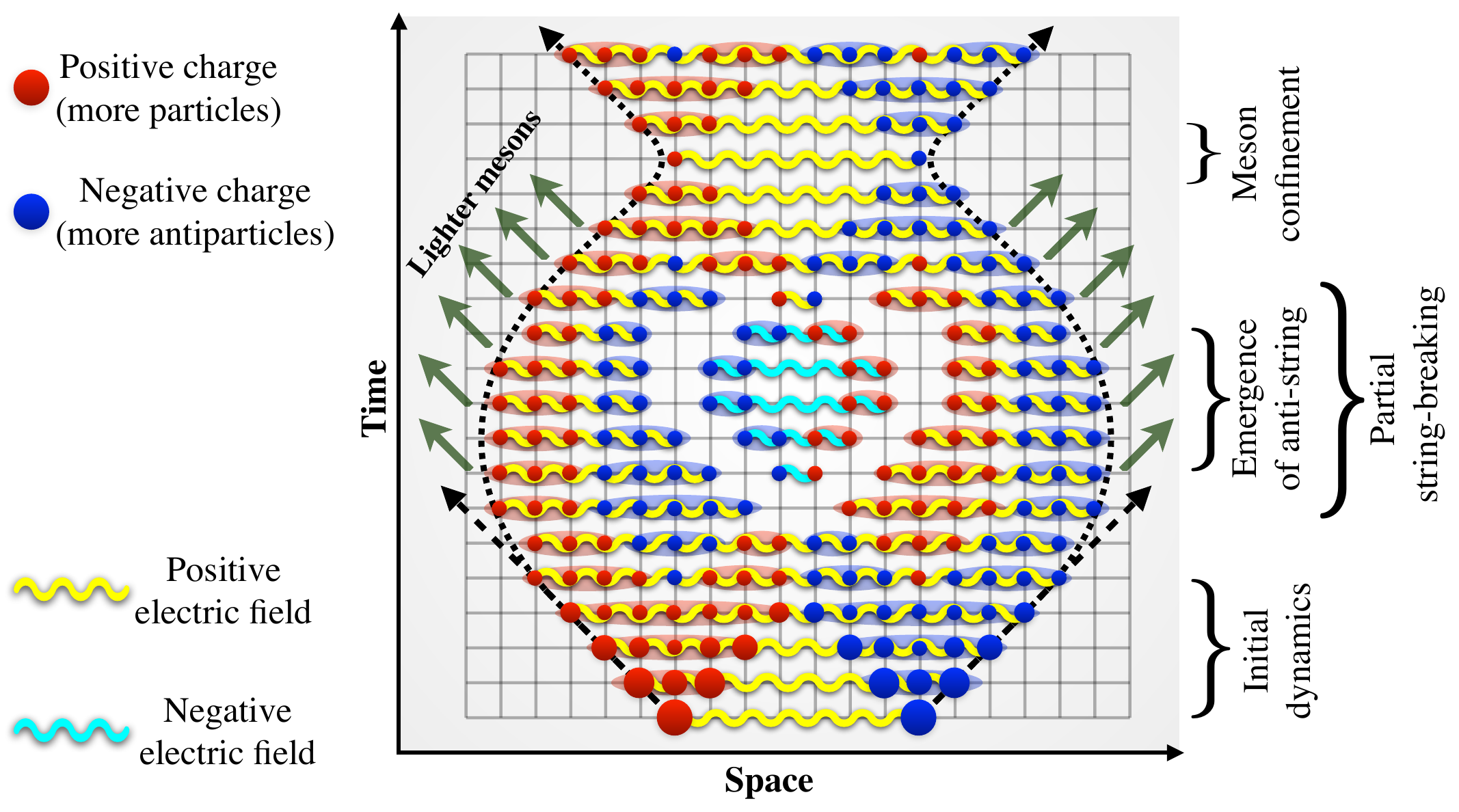}
\caption{(Color online.) Semi-classical sketch of the confining dynamics of BSM. We prepare a well separated pair of particle and antiparticle connected by an electric-flux tube. Initially they  start spreading as if they were free, however their trajectories bend due to the energetic cost of creating larger electric-flux tubes. New dynamical charges are also created during the evolution and  partially screen the electric field.
Still the electric field oscillates coherently and can form an anti-string, creating a central core of strongly correlated bosons.
The density of bosons in the core can get depleted through the radiation of lighter mesons that freely propagate.
\label{fig:schem2}}
\end{figure}

The Hamiltonian version of the BSM is made by two bosonic species that behave, in the non-interacting regime, as a discretized version of the Klein-Gordon (KG) field theory \cite{kogut_prd_1975, creutz_prd_1977, creutz_prd_1979, article_version}. Both species couple to the  $U(1)$  field, one representing the particle and the other the anti-particle.
The low energy spectrum of this system is always gapped even for massless bosons, and as a result the bosons are always confined \cite{supple, article_version}.
By using state-of-the-art matrix-product states (MPS) techniques \cite{schollwock_aop_2011, verstraete_arxiv_2004, orus_aop_2014, bridgeman_jphysa_2017, ran_arxiv_2017, crosswhite_prb_2008, hauke_njp_2010, koffel_prl_2012,  hauke_prl_2013, milsted_prd_2013, zalatel_prb_2015, banuls_prx_2017, paeckel_arxiv_2019} in their gauge-invariant version \cite{byrnes_prd_2002, sugihara_jhep_2005,tagliacozzo_prb_2011,banuls_jhep_2013,tagliacozzo_prx_2014,silvi_njp_2014, kuhn_prl_2015, haegeman_prx_2015}, we analyze the OED in both the massless and the massive regimes of the theory.

We start the OED by creating a particle-antiparticle pair separated by a  distance and observe the processes sketched in  Fig. \ref{fig:schem2}. The two bosons are connected by an electric-flux tube that bends their trajectories  inwards. The process is closer to that observed for the Ising model
 \cite{kormos_natphys_2017} than to what has been observed in 
the FSM \cite{pichler_prx_2016, kuhn_prl_2015}.

The initial bosons form a core of a strongly correlated matter that oscillates several times around its original position. Even though, in the small mass regime, the matter density in the core is gradually depleted by the  radiation of free lighter mesons, the core stays quantitatively different from the rest of the system, thus keeps a strong memory of the initial state and evades thermalization. The lack of total screening of the electric field by the dynamically generated matter constitutes the main origin of this phenomenon.  

However, similar phenomena have been observed  in very different contexts such as  holographic quenches \cite{craps_jhep_2015,abajo-arrastia2014,dasilva2015,dasilva_jhep_2016,myers_jphysa_2017}, spins systems \cite{brenes_prl_2018, robinson_prb_2019, james_prl_2019}, large $N$ gauge theories \cite{cubero_arxiv_2019} and quantum link models \cite{surace_arxiv_2019} and Rydberg atoms \cite{turner_natphys_2018, sala_arxiv_2019}.  We thus provide a generic framework to understand such phenomena in terms of  entropy production \cite{rigol_nature_2008, vidmar_jstatmech_2016}.

In particular,  the lack of homogeneity in space of the  classical part of the entanglement entropy provides a generic footprint  that allows one to understand the origin of the slow dynamics we observe here and its connection with similar phenomena observed elsewhere.

\paragraph{Model.--}

The BSM Lagrangian density is   given by 
$\mathcal{L} = -\big[D_{\mu} \phi\big]^* D^{\mu} \phi -m^2 |\phi|^2 - \frac{1}{4}F_{\mu\nu}F^{\mu\nu}$ \cite{peskin_book},
where $\phi$ is the complex  scalar field,
$D_{\mu} = (\partial_{\mu} + i q A_{\mu})$ is the covariant derivative with $q$ and $A_{\mu}$ being the electronic-charge and vector potential respectively, $m$ is the bare mass of the particles, and $F_{\mu\nu}$ is the electromagnetic field tensor.  In 1D,  after fixing the temporal gauge, $A_t = 0$, we get the discretized Hamiltonian (in dimensionless units) as \cite{supple,article_version}
\small
\begin{eqnarray}
\hat{H} &=& \sum_j \hat{L}_j^2 + 
2 \left(x \left(\left(m/q\right)^2 + 2 x\right)\right)^{1/2} \ \sum_j \big(\hat{a}_j^{\dagger}\hat{a}_j + \hat{b}_j\hat{b}_j^{\dagger}\big) \nonumber \\
&-& \frac{x^{3/2}} {\left(\left(m/q\right)^2 +2 x\right)^{1/2}} \sum_j \left[ \big(\hat{a}^{\dagger}_{j+1} + \hat{b}_{j+1}\big) \hat{U}_j \big(\hat{a}_{j} + \hat{b}^{\dagger}_{j}\big) + \text{h.c.}\right], \nonumber \\
\label{eq:Hamil1}
\end{eqnarray} 
\normalsize
where $\{\hat{a}_j^{\dagger}, \hat{a}_j\}$, $\{\hat{b}_j^{\dagger}, \hat{b}_j\}$ are the bosonic creation-annihilation operators corresponding to charged particles and antiparticles respectively, $\hat{L}_j$ is the electric field operator residing on the bond between sites $j$ and $j+1$  with $\{\hat{U}_j, \hat{U}_j^{\dagger}\}$ being 
$U(1)$ ladder operators satisfying $[\hat{L}_j, \hat{U}_l] = -\hat{U}_j \delta_{jl}$ and $\ [\hat{L}_j, \hat{U}^{\dagger}_l] = \hat{U}^{\dagger}_j \delta_{jl}$, and $x = 1/(a^2 q^2)$ with $a$ being the lattice-spacing. 

The Hamiltonian is invariant under local $U(1)$ transformations: $\hat{a}_j \rightarrow e^{i \alpha_j} \ \hat{a}_j$,
$\hat{b}_j \rightarrow e^{-i \alpha_j} \ \hat{b}_j$,
$\hat{U}_j \rightarrow e^{-i \alpha_j} \ \hat{U}_j \ e^{i \alpha_{j+1}}$, where the corresponding Gauss law generators are given by
$\hat{G}_j = \hat{L}_j - \hat{L}_{j-1} - \hat{Q}_j$,
with  $\hat{Q}_j = \hat{a}_j^{\dagger} \hat{a}_j - \hat{b}_j^{\dagger} \hat{b}_j$ being the dynamical charge \cite{supple}. The physical subspace is spanned by the set of states, $\ket{\Psi}$, that are annihilated by $\hat{G}_j$, i.e., $\hat{G}_j \ket{\Psi} = 0 \  \forall j$. Using this conservation law, in an open chain, one can integrate-out the gauge fields  using the transformation, $\hat{L}_j = \sum_{l \leqslant j} \hat{Q}_l$, where we consider the static electric field on the left of the chain to be zero.
The effective Hamiltonian for matter fields, once we have integrated out the photons, contains long-range intra-species repulsion and inter-species attraction.

The continuum limit of the system is achieved by taking the two limits: 
$N \rightarrow \infty$ and $x \rightarrow \infty$. However, since our study is motivated by the 
near future experimental realizations with cold-atoms, we consider an  open chain made of a finite number of lattice 
sites $N = 60$ at a fixed value of $x$ that without loss of generality we consider $x = 2$.
We obtain systematic matrix-product state approximations to the  the ground-state of the system using the density matrix renormalization group (DMRG) \cite{white_prl_1992, white_prb_1993, schollwock_rmp_2005, schollwock_aop_2011}. This is the starting point  
for the  time-evolution, that we approximate by  using the time-dependent variational principle (TDVP) \cite{haegeman_prl_2011, koffel_prl_2012, haegeman_prb_2016,paeckel_arxiv_2019,hauke_prl_2013} (see \cite{supple} for more details).

\emph{Real-time evolution.--}
We start the OED of the BSM by creating  two extra dynamical charges of opposite signs on the top of the ground state $\ket{\Omega}$  of  the Hamiltonian  \eqref{eq:Hamil1}. They are located at positions $N/2 - R$ and $N/2+R+1$ respectively, and are connected by a string of electric field of length $2 R+ 1$. They are created by means of the non-local operator
$\hat{M}_R$, defined as 
\small
\begin{equation}
 \hat{M}_R \equiv \left(\hat{a}_{\frac{N}{2} - R}^{\dagger} + \hat{b}_{\frac{N}{2} - R}\right) \left[\prod_{j = \frac{N}{2} - R}^{\frac{N}{2} + R} \hat{U}_{j}^{\dagger}\right]  \left(\hat{a}_{\frac{N}{2} + R + 1} + \hat{b}_{\frac{N}{2} + R +1}^{\dagger}\right).
 \label{eq:MR}
 \end{equation}
\normalsize
As a result, the initial state is $\ket{\psi(t=0)} = \mathcal{N} \hat{M}_R \ket{\Omega}$, where $\mathcal{N}$ is a normalization constant. The state  has the extra energy $\approx (2 R + 1) + 4 \left(x ((m/q)^2 + 2 x)\right)^{1/2}$ 
above the ground state energy. 
When $R$ is a finite fraction of the system size, the initial state has an extensive extra energy.  As a result, the evolution of $\ket{\psi(t=0)}$ under \eqref{eq:Hamil1} should mostly be driven by high-energy excited states.

\emph{Confinement and string-breaking.--}
We would expect the charges, initially created at distance $2R+1$, to separate by stretching the electric-flux string up to a critical distance that only depends on the boson mass. There ultimately the string would break as a result of boson pair-production from the vacuum. After that,  
lighter mesons would propagate freely %
(similarly to the FSM case \cite{pichler_prx_2016}).
However, in the BSM we do not observe such a scenario. We only see a \emph{partial} screening of the initial electric-field string, leading to a \emph{partial} string-breaking, even for massless bosons. Our observations can be explained using a semi-classical cartoon presented in  Fig.~\ref{fig:schem2} displaying four main features of the OED:

\begin{itemize}[leftmargin=*, label={}]
\item \emph{Confinement and slow dynamics induced by the partial screening of electric field:} The electric-flux string creates a long-lived metastable state in the middle of the system that oscillates and  contract inwards \cite{craps_jhep_2015, dasilva_jhep_2016, myers_jphysa_2017, robinson_prb_2019}, reducing the velocity of charges and  bending the initial light cone \cite{kormos_natphys_2017}. The rate of production of new bosons is not sufficient to completely screen the electric field, so that their light-cone bends as the charges  slow-down  and eventually start  refocusing. The two charges are thus confined again into an extended meson that wobbles at a well defined frequency (Fig.~\ref{fig:schem2}).

\item \emph{String-inversion in the bulk:} The string does not break, not even for massless bosons, but rather undergoes at least a pair of coherent oscillations (see also \cite{surace_arxiv_2019}). For lighter masses, we observe a string-inversion phenomenon and slow damping of it by the radiation of lighter mesons.  

\item \emph{Mesons radiation, the two domains:} While in the bulk the string contracts and expands periodically, at the boundaries the electric flux is partially screened and the string can break into pieces forming lighter mesons. They are free to escape the confined region and fly away with a constant velocity, creating two separate domains. We observe a core region where the bosons are confined (a confined domain), and an outer region where they escape in the form of lighter mesons (a deconfined domain). 
The radiation of lighter mesons slowly depletes the electric field and the cloud of bosons in the confined region.

\end{itemize}

\begin{figure}
\includegraphics[width=\linewidth]{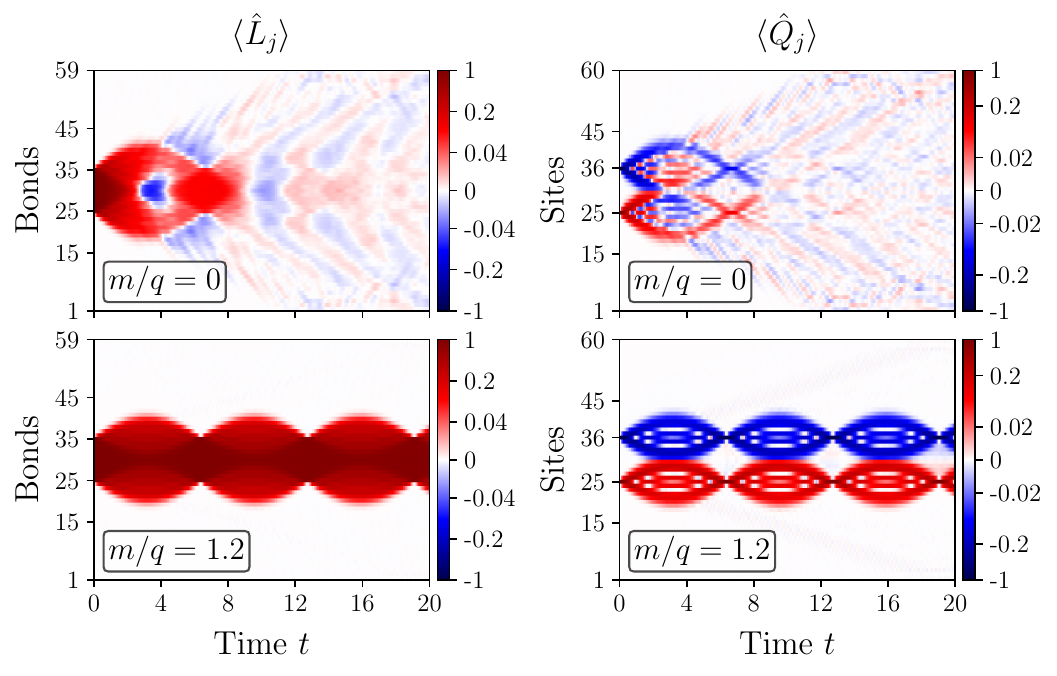}
\caption{(Color online.) Dynamics of the electric field $\hat{L}_j$ and the dynamical charge $\hat{Q}_j$ for $R=5$. 
}
\label{fig:dyn1}
\end{figure}

\paragraph{Quantitative results about the lack of thermalization.--}
The cartoon in Fig.~\ref{fig:schem2} tries to summarize the numerical  results  presented  in Fig.~\ref{fig:dyn1}. There we plot both  the electric field ($\hat{L}_j$) on the left   and the dynamical charges ($\hat{Q}_j$) on the right for two paradigmatic values  $m/q = 0$ and $1.2$ that together display all the phenomena we have listed previously. The dynamics is generated by acting with the operator $\hat{M}_{R=5}$  (Eq. \eqref{eq:MR})  on the vacuum of \eqref{eq:Hamil1}. In the massless scenario (top row of Fig.~\ref{fig:dyn1}), we appreciate the \emph{partial screening} of the electric field, visible in the bending of the bosons' light cone; the  appearance of an \emph{anti-string} in the bulk of the system due to the \emph{string-inversion} at $t\simeq 2.5$; the \emph{meson radiation} from the edges of the confined bulk  starting around  $t \gtrsim 4$. We also see that the confined core of bosons is gradually depleted and disappears for times around $t\simeq 10$.

As we increase the mass, we slowly move towards the heavy-bosons scenario of  $m/q = 1.2$ (bottom row of Fig. ~\ref{fig:dyn1}), where the radiation of free mesons is strongly suppressed and the confined core is basically surrounded by the vacuum.  We indeed observe an almost perfect periodic oscillation of the electric string. Intermediate regimes (not shown, see \cite{supple}) display interesting features. For example  for $m/q = 0.25$, the anti-string, visible in $m/q = 0$ case, disappears  and there is no string-breaking  for $m/q \gtrsim 0.5$. 

While for  $m/q=0$  the concentration of dynamical charges in the confined domain gradually disappears for $t \gtrsim 10$, it persists in the asymptotic states for larger values of the mass. However, the confined core still lingers on for a long time, even in the massless case, which can be perceived through the fluctuation of  $\hat{L}_j$ or  $\hat{Q}_j$, or through the spreading of the entanglement entropy.

From the above discussions it is clear that confinement strongly hinders the thermalization process in the system and the `equilibrated' states are far from thermal. 

\begin{figure}
\includegraphics[width=\linewidth]{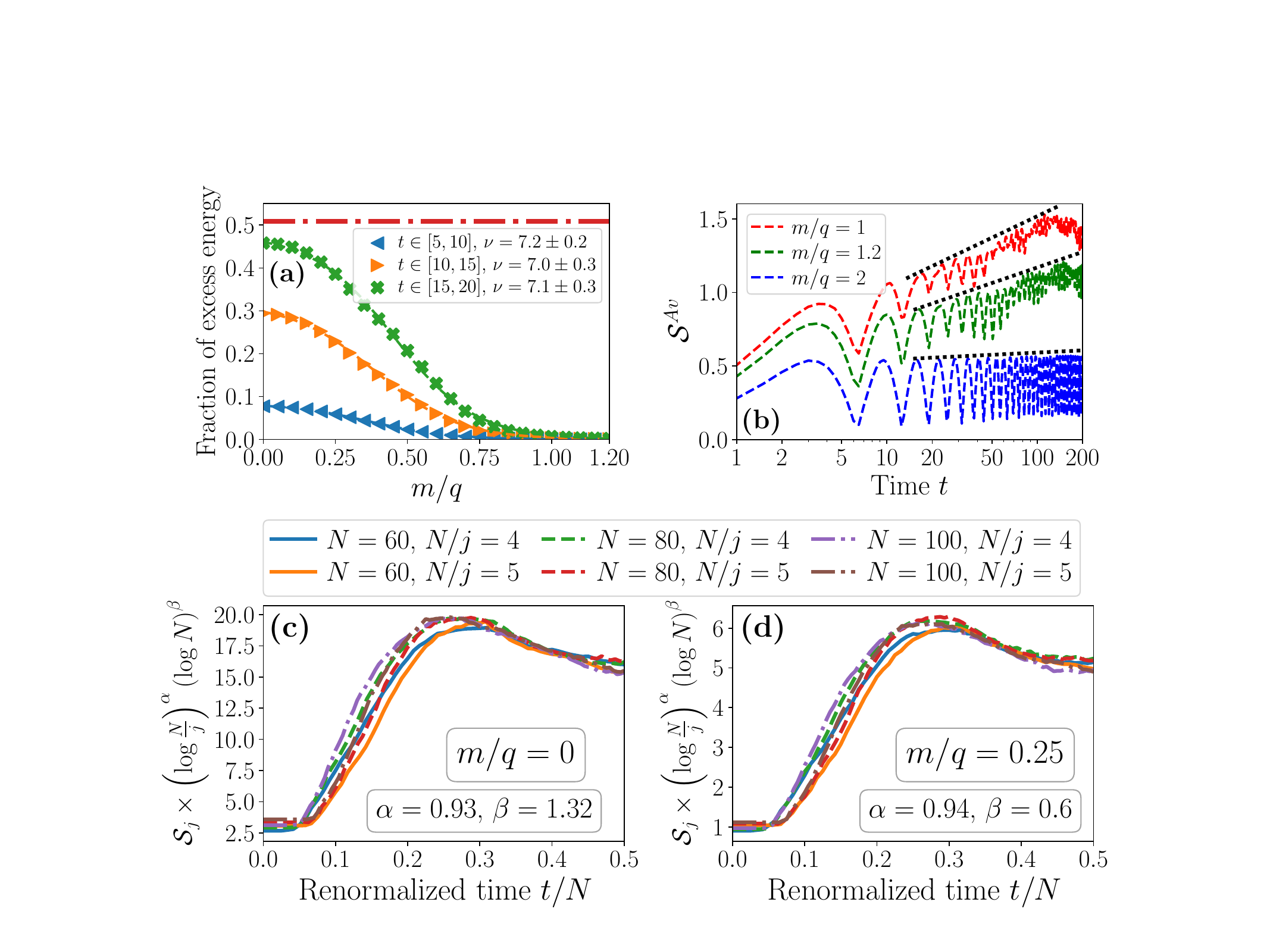}
\caption{(Color online.) (a) Time-averaged fraction of the initial extra energy that leaks from the core of the system towards the boundaries for different ratios of $m/q$ and different time-windows. We measure the excess energy across the bonds $[1, 15] \cup [45, 59]$ to obtain the fraction. The red dot-dashed line marks the ideal homogeneous value. The time-averaged fraction has been fitted by a sigmoid curve with exponent $\nu$ (see text).
(b) Long time behavior of the growth of entanglement entropy in the central region for larger values of masses.
(c)-(d) Pattern of entanglement entropy in the deconfined domain for different system and bipartition sizes for $m/q = 0$ (c) and $0.25$ (d).}
\label{fig:comb}
\end{figure}

Despite the qualitative difference in Fig. \ref{fig:dyn1} between the light and heavy mass regime, we show now that they are both non-thermal and continuously connected. In Fig. \ref{fig:comb}(a), we consider the time-averaged fraction of the initial excess energy that leaks to the outer regions of the system. If the system eventually thermalizes the energy should spread uniformly (beside finite size boundary effects). 
In Fig \ref{fig:comb}(a), the uniform value is represented by a red line, and different colors and symbols encode different time windows. The plot shows that for any time and value of $m/q$ the energy stays trapped in the bulk as witnessed by the lower amount of energy leakage  to the edges of the system  than what is expected in an homogeneous system. Furthermore, the leakage  continuously decreases as we increase $m/q$ in a smooth way, without any transitions as shown by the perfect fit of our numerical data to the sigmoid curve 
$\sim 1/\left( 1+ \exp\left(\nu m/q - c \right)\right)$ with exponent $\nu$. For $m/q$ larger than $1$ the leaking is practically zero.
There is thus a single dynamics regime characterized by a strong memory of the initial state. 

The best way to characterize the slow dynamics is by considering the growth and spreading of entanglement. The entanglement entropy, $\mathcal{S}_j$, measured across the bond between the sites $j$ and $j+1$ is defined as 
$\mathcal{S}_j(t) = - \mbox{Tr} \left[\rho_j(t) \ln \rho_j(t) \right]$, with $\rho_j(t) = \mbox{Tr}_{j+1, j+2,..,N} \ket{\psi(t)}\bra{\psi(t)}$ being the reduced density matrix to the left of the $j^{th}$ bond. The spread of entanglement out of equilibrium has been discussed originally in \cite{calabrese_jstatmech_2005} 
and more recently in the context of generalized thermalization 
\cite{rigol_nature_2008, vidmar_jstatmech_2016}. {In the BSM, the spread of entanglement is strongly modified by the effect of confinement similar to the evolution of local observables mentioned earlier (see \cite{supple}).}

We start analyzing the growth of entanglement entropy with time for larger masses  as the simulations are less demanding \cite{note_large_mass}. By plotting the long time evolution of the average entropy in the central region, $\mathcal{S}^{Av} = \frac{1}{2 R + 1} \sum_{j = N/2 - R}^{N/2 + R} \mathcal{S}_j$, we can appreciate how the entropy strongly oscillates with an envelope that grows logarithmically with time before saturating at $t \simeq 160$ (Fig. \ref{fig:comb}(b)). Such a logarithmic growth of entropy bears a strong resemblance to observations in  many-body-localized systems (MBL) \cite{znidaric_prb_2008, bardarson_prl_2012}. Unlike in MBL, the saturated value of entanglement entropy does not depend on the system size (see \cite{supple}).

For lighter masses we cannot achieve such long times, but we can still analyze the scaling of the entropy with the system size in the asymptotic states we obtain.  We consider $N = 60, 80,$ and $100$ and $R = N/ 10$, so that the length of the initial string grows proportional to the system size, thus providing an extensive amount of energy on the top of the ground state and fulfilling one of the conditions for thermalization. In a thermalized system, the entropy of a region grows proportional to its size (volume-law), while in the non-ergodic scenario the entropy increases slower  than that.
We expect that the deconfined domain slowly becomes thermal, due to the radiation of lighter mesons, while the confined domain remains ``non-thermal'' showing coherent oscillations.

Indeed in the confined domain the entropy shows a perfect area-law (see \cite{supple}). Neither its maximum value nor its average over the confined domain depend on the size of the region as shown in \cite{supple}.

On the other hand, from the scaling plots (Figs. \ref{fig:comb}(c)-(d)), we find that the entropy in the deconfined region varies as  
$\mathcal{S}_j \propto \left(\log\frac{N}{j}\right)^{-\alpha} \left(\log N\right)^{-\beta}$,
where the exponents $\alpha \approx 1$ and $\beta$ depend on $m/q$. For fixed $N$, the entropy increases sub-linearly for small bipartitions, then turns into a linear increase for intermediate distances, and ultimately shows faster than volume-law growth before saturating into the confined domain. From this scaling form, it seems reasonable to expect that  the deconfined region would actually `thermalize' at intermediate distances, far away from both the core and the boundaries of the system. This complicated scaling form seems to match the behavior expected in a thermalized region with the presence of a physical boundary and of a confined core.

\paragraph{Classical and distillable entanglement dynamics.--}

\begin{figure}
\includegraphics[width=\linewidth]{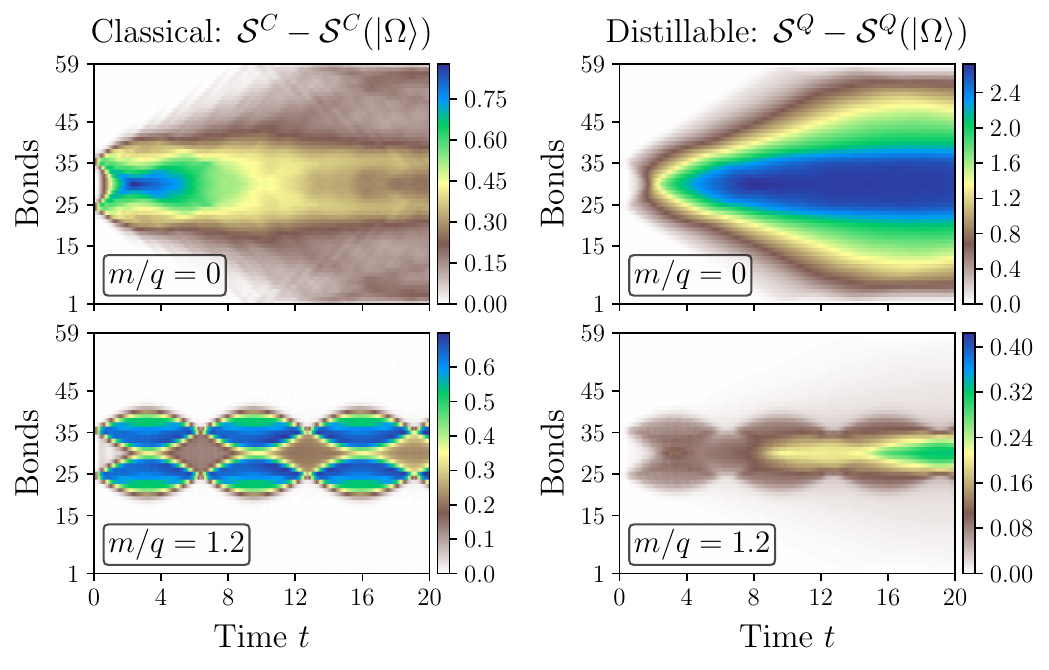}
\caption{(Color online.) Time-evolution of the classical part $\mathcal{S}^C$ of entanglement entropy (left column) and the distillable entanglement entropy $\mathcal{S}^Q$ (right column). Here we subtract the corresponding entropies of the ground state $\ket{\Omega}$ and consider the case of $R=5$.}
\label{fig:entQC}
\end{figure}

As part of  the gauge symmetry, the system possesses a global $U(1)$ symmetry corresponding to the conservation of total dynamical charge, $\sum_j \hat{Q}_j$. As a result,  any reduced density matrix is block-diagonal in  different charge sectors:
$\rho = \bigoplus_{{Q}} \tilde{\rho}_{_Q} = \bigoplus_{Q} p_{_Q} \ \rho_{_Q}$,
where $p_{_Q} = \mbox{Tr} \left[\tilde{\rho}_{_Q}\right]$ and $\rho_{_Q} = \tilde{\rho}_{_Q} / p_{_Q}$. Therefore, following  \cite{schuch_prl_2004, schuch_pra_2004, donnelly_prd_2012, radicevic_arxiv_2014, casini_prd_2014, casini_arxiv_2019,Lukin_science_2019,Sierant_scipost_2019} the entanglement entropy can be divided into two parts as
$ \mathcal{S}(\rho) = -\sum_Q p_{_Q} \ln p_{_Q} + \sum_Q p_{_Q} \mathcal{S}(\rho_{_Q})$,
where the first term is the classical part ($\mathcal{S}^C$) -- the Shannon entropy between different quantum blocks indicated by $Q \in [.., -1, 0, 1,...]$, and the second term, the distillable entanglement ($\mathcal{S}^Q$) that has a clear operational significance \cite{schuch_prl_2004}.

Fig.~\ref{fig:entQC} shows the time-evolution of both parts of the entanglement entropy. The classical part clearly demarcates confined and deconfined domains by remaining \emph{sharply} concentrated only in the confined domain. 
Most of the bosons are confined in the core and thus populate several superselection sectors as witnessed by the large value of the classical entropy. In the deconfined domain, the density of lighter mesons that escape from the core is low, meaning that the only populated superselection sectors are  $Q = 0, \pm 1$.  Furthermore, since mesons are short and dilute (on average we have less than one meson per site),  $p_{_{Q=\pm 1}} \ll p_{_{Q=0}}$. We thus conclude that higher charge sectors are off-resonant in this domain, as they would  require a higher density of mesons. 

The distillable part of the entropy spreads throughout the system in a `more ergodic' manner. It still retains a distinctly higher value in the confined domain within this time-scale, but we believe it can be a finite time effect. Summarizing, the strong spatial inhomogeneity of the time-evolved entropies, especially of their classical part, gives us a clear generic signature of the lack-of-thermalization in the system.

\paragraph{Conclusion.--}
We have shown that the out-of-equilibrium dynamics generated by creating a pair of bosons on the top of the  interacting vacuum of a  gauge theory is strongly affected by confinement. The asymptotic states generated are highly inhomogeneous.
They are 
made by a confined core that displays long-lived oscillations and entropically fulfills the area-law, surrounded either by the vacuum (for heavy bosons) or by an almost thermal cloud of light mesons (for lighter bosons). In  both cases, the long-time asymptotic states  have a strong memory of the initial state and thus evade thermalization in its simplest form. 
The phenomena observed are
reminiscent of the observations of the presence of highly non-thermal states in the quantum Ising model  \cite{james_prl_2019}. 
In a forthcoming paper \cite{article_version} we will relate 
our findings to  the intricate phase diagram of the BSM at equilibrium. 

Interestingly, the initial confined core persists up to a  very long time, as witnessed by the entanglement entropy. Indeed the best signature of this phenomenon is given by the spatial profile of the classical part of the entanglement entropy.  We have pushed the current algorithms to their limit, and larger system size and times will have to  be addressed either using the next-generation of tensor-network methods (for recent proposal see \cite{surace_prb_2019} and references therein) or experimentally by quantum simulators \cite{mil_arxiv_2019}. 

\begin{acknowledgments}
We thank Subhroneel Chakrabarti, Sanjukta Kundu, Marek M. Rams, Krzysztof Sacha, Piotr Sierant, Piotr Korcyl, Krzysztof Biedro{\'n}, Gert Aarts, Alessio Celi, Valentin Kasper,  E. Miles Stoundenmire and Matthew Fishman for useful discussions.  MPS algorithms have been implemented using ITensor library v2 (\url{https://itensor.org}). 
T. C. and J. Z. acknowledge support by PL-Grid Infrastructure and by 
National Science Centre (Poland) under   QuantERA QTFLAG
project  2017/25/Z/ST2/03029.
M.L. acknowledges support by the Spanish Ministry MINECO (National Plan
15 Grant: FISICATEAMO No. FIS2016-79508-P, SEVERO OCHOA No. SEV-2015-0522, FPI), European Social Fund, Fundaci{\'o} Cellex, Generalitat de Catalunya (AGAUR Grant No. 2017 SGR 1341 and CERCA/Program), ERC AdG OSYRIS and NOQIA, and the National Science Centre, Poland-Symfonia Grant No. 2016/20/W/ST4/00314. 
LT is supported by the MINECO  RYC-2016-20594 fellowship and the MINECO PGC2018-095862-B-C22 grant, and acknowledges the support from the European Union Regional Development Fund within the ERDF Operational Program of Catalunya, Spain (project QUASICAT/QuantumCat, ref. 001- P-001644).
Upon submitting this paper to the arXiv, we became aware of a complementary approach that tries to explain the slow thermalization and the lack of thermalization observed in similar models in terms of fractons \cite{pai_arxiv_2019}. It will be interesting to understand if that approach can also be used here.
\end{acknowledgments}

\bibliography{main_bsm_lt_jz.bbl}


\end{document}


\title{Supplementary matarial to ``Confinement and lack of thermalization after quenches in bosonic Schwinger model"}

\author{Titas Chanda}
\email{titas.chanda@uj.edu.pl}
\affiliation{Instytut Fizyki im. Mariana Smoluchowskiego, Uniwersytet Jagiello\'nski, \L{}ojasiewicza 11, 30-348 Krak\'ow, Poland} 

\author{Jakub Zakrzewski}
\affiliation{Instytut Fizyki im. Mariana Smoluchowskiego, Uniwersytet Jagiello\'nski, \L{}ojasiewicza 11, 30-348 Krak\'ow, Poland} 
\affiliation{Mark Kac Complex Systems Research Center,  Jagiellonian University in Krakow, 30-348 Krak{\'o}w, Poland}

\author{Maciej Lewenstein}
\affiliation{ICFO-Institut de Ci\`encies Fot\`oniques, The Barcelona Institute of Science and Technology, Av. Carl Friedrich
Gauss 3, 08860 Castelldefels (Barcelona), Spain}
\affiliation{ICREA, Passeig Lluis Companys 23, 08010 Barcelona, Spain}

\author{Luca Tagliacozzo}
\affiliation{Department of Physics and SUPA, University of Strathclyde, Glasgow G4 0NG, UK}
\affiliation{Departament de F\'{\i}sica Qu\`antica i Astrof\'{\i}sica and Institut de Ci\`encies del Cosmos (ICCUB), Universitat de Barcelona,  Mart\'{\i} i Franqu\`es 1, 08028 Barcelona, Catalonia, Spain}

\maketitle

\subsection{Derivation of the Hamiltonian}

The Lagrangian density for the BSM \cite{peskin_book} is given by
\begin{equation}
\mathcal{L} = -\left[ D_{\mu} \phi \right]^* D^{\mu} \phi -m^2 |\phi|^2 - \frac{1}{4}F_{\mu\nu}F^{\mu\nu},
\end{equation}
where $\phi$ is the complex scalar field, $D_{\mu} = (\partial_{\mu} + i q A_{\mu})$ is the covariant derivative with $q$ and $A_{\mu}$ being the electronic charge and electromagnetic vector potential respectively, $m$ is the mass of the particles, and $F_{\mu\nu}$ is the electromagnetic field tensor. Here, we use the metric convection $(-1,1,1,1)$ or $(-1,1)$ (in 1+1 dimensions).  In 1+1 dimensions,  after fixing the temporal gauge $A_t(x, t) = 0$, we get the quantum Hamiltonian as 
\begin{eqnarray}
\hat{H} &=& \int dx \bigg[\frac{1}{2} \hat{E}_x^2(x) +  \hat{\Pi}^{\dagger}(x) \hat{\Pi}(x) +  m^2 \hat{\phi}^{\dagger}(x) \hat{\phi}(x)  \nonumber \\
&+& \left(\partial_x - i q \hat{A}_x(x)\right) \hat{\phi}^{\dagger}(x) \left(\partial_x + i q \hat{A}_x(x)\right) \hat{\phi}(x)
\bigg],
\end{eqnarray}
where $\hat{E}_x(x)$, $\hat{\Pi}(x)$, and $\hat{\Pi}^{\dagger}(x)$ are the canonical conjugate operators corresponding to $\hat{A}_x(x)$, $\hat{\phi}(x)$, and $\hat{\phi}^{\dagger}(x)$ respectively,
satisfying $[\hat{A}_x(x_1), \hat{E}_x(x_2)] = [\hat{\phi}(x_1), \hat{\Pi}(x_2)] = [\hat{\phi}^{\dagger}(x_1), \hat{\Pi}^{\dagger}(x_2)] = i \delta(x_1-x_2)$. 
We can discretize this Hamiltonian on a 1D lattice having lattice-spacing $a$ in a straightforward way such that the matter fields $\{\hat{\phi}_j,  \hat{\phi}_j^{\dagger},
\hat{\Pi}_j,  \hat{\Pi}_j^{\dagger}\}$ reside on lattice site $j$, while the gauge fields $\{\hat{A}_j,  \hat{E}_j\}$ act on the bonds between lattice points, e.g. $j$ and $j+1$.
The Hamiltonian, thus discretized, reads as
\begin{eqnarray}
\hat{H} &=& \frac{a}{2} \sum_j \hat{E}_j^2 + \frac{1}{a} \sum_j \hat{\Pi}_j^{\dagger} \hat{\Pi}_j + \left(a m^2 + \frac{2}{a} \right) \sum_j \hat{\phi}_j^{\dagger} \hat{\phi}_j \nonumber \\
&-& \frac{1}{a} \sum_j \left[ \hat{\phi}_{j+1}^{\dagger} \exp(-i q \hat{A}_j) \hat{\phi_j} + \mbox{ h.c.} \right],
\end{eqnarray} 
where the operators have been rescaled to satisfy the commutation relations $[\hat{A}_j, \hat{E}_k] =  \ [\hat{\phi}_j, \hat{\Pi}_k] = [\hat{\phi}^{\dagger}_j, \hat{\Pi}^{\dagger}_k] = i \delta_{jk}$. 
In the next few steps, we introduce bosonic operators $\hat{a}_j$ and $\hat{b}_j$ as
\begin{eqnarray}
\hat{\phi}_j = \frac{1}{\sqrt{2}} \left(\hat{a}_j + \hat{b}_j^{\dagger}\right), \ \hat{\Pi}_j = \frac{i}{\sqrt{2}} \left(\hat{a}^{\dagger}_j - \hat{b}_j\right), \nonumber \\
\hat{\phi}^{\dagger}_j = \frac{1}{\sqrt{2}} \left(\hat{a}^{\dagger}_j + \hat{b}_j\right), \ \hat{\Pi}^{\dagger}_j = \frac{i}{\sqrt{2}} \left(\hat{b}^{\dagger}_j - \hat{a}_j\right),
\end{eqnarray}
rescale the gauge fields as $\hat{L}_j = \hat{E}_j/q, \ \hat{\theta}_j = q \hat{A}_j$, and multiply the Hamiltonian by $1/aq^2$ to make it dimensionless. 
The Hamiltonian now becomes 
 \begin{eqnarray}
\hat{H} &=& \sum_j \hat{L}_j^2 + \left((m/q)^2 + 3x\right) \sum_j \big(\hat{a}_j^{\dagger}\hat{a}_j + \hat{b}_j\hat{b}_j^{\dagger}\big) \nonumber \\
&+& \left((m/q)^2 + x\right) \sum_j \big(\hat{a}_j^{\dagger}\hat{b}_j^{\dagger} + \hat{a}_j\hat{b}_j\big) \nonumber \\
&-& x \sum_j \left[ \big(\hat{a}^{\dagger}_{j+1} + \hat{b}_{j+1} \big) \hat{U}_j  \big(\hat{a}_{j} + \hat{b}^{\dagger}_{j} \big) + \text{h.c.}\right],
\label{eq:Hamil3}
\end{eqnarray} 
where $\hat{U}_j = \exp(-i \hat{\theta_j})$ and $\hat{U}^{\dagger}_j = \exp(i \hat{\theta_j})$ are the ladder operators satisfying
$[\hat{L}_j, \hat{U}_j] = -\hat{U}_j$ and $\ [\hat{L}_j, \hat{U}^{\dagger}_j] = \hat{U}^{\dagger}_j$ respectively, and  $x = 1/a^2 q^2$. To further simplify the Hamiltonian, we employ a local Bogoliubov transformation as
\begin{eqnarray}
\hat{a}_j &\rightarrow & \cosh(\theta_j) \  \hat{a}_j + \sinh(\theta_j) \ \hat{b}_j^{\dagger}, \nonumber \\
\hat{b}_j &\rightarrow & \cosh(\theta_j) \ \hat{b}_j + \sinh(\theta_j) \ \hat{a}_j^{\dagger},
\end{eqnarray}
where $\theta_j = -\frac{1}{2}\tanh^{-1} \left[ \frac{\left(m/q\right)^2 + x} {\left(m/q\right)^2 + 3x} \right]$ for all $j$, such that our final Hamiltonian is given as
\begin{widetext}
\begin{equation}
\hat{H} = \sum_j \hat{L}_j^2 + 
2 \left(x \left(\left(m/q\right)^2 + 2 x\right)\right)^{1/2} \ \sum_j \big(\hat{a}_j^{\dagger}\hat{a}_j + \hat{b}_j\hat{b}_j^{\dagger}\big) 
-\frac{x^{3/2}} {\left(\left(m/q\right)^2 +2 x\right)^{1/2}} \sum_j \left[ \big(\hat{a}^{\dagger}_{j+1} + \hat{b}_{j+1}\big) \hat{U}_j \big(\hat{a}_{j} + \hat{b}^{\dagger}_{j}\big) + \text{h.c.}\right]. 
\label{eq:Hamil4}
\end{equation}
\end{widetext}
We refer the bosons `$a$' and `$b$' as particles and antiparticles respectively.

\subsection{Local gauge invariance and Gauss law generators}
It can be straightforwardly verified that 
the Hamiltonian in Eq.~\eqref{eq:Hamil4} is invariant under local $U(1)$ gauge transformations:
\begin{eqnarray}
&&\hat{a}_j  \rightarrow e^{i \alpha_j} \ \hat{a}_j, \
\hat{b}_j  \rightarrow e^{-i \alpha_j} \ \hat{b}_j, \\
&&\hat{U}_j  \rightarrow e^{-i \alpha_j} \ \hat{U}_j \ e^{i \alpha_{j+1}}.
\end{eqnarray}
 Corresponding Gauss law generators can be obtained from the Euler-Lagrange equation of the field $A_t$, i.e.,
\begin{eqnarray}
&&\left[ \partial_{\mu}\left[\frac{\partial \mathcal{L}}{\partial (\partial_{\mu} A_t)} \right] - \frac{\partial \mathcal{L}}{\partial A_t}\right]_{A_t \rightarrow 0} = 0, \nonumber \\
&&\Rightarrow \partial_x E_x(x) = i q \left(\phi^*(x) \Pi^*(x) - \phi(x) \Pi(x) \right) ,
\end{eqnarray}
which, after quantization with normal ordering and discretization similar to the Hamiltonian, gives us the Gauss law generators as
\begin{eqnarray}
\hat{G}_j = \hat{L}_j - \hat{L}_{j-1} - \underbrace{\left(\hat{a}_j^{\dagger} \hat{a}_j - \hat{b}_j^{\dagger} \hat{b}_j \right)}_{\hat{Q}_j},
\end{eqnarray}
where the dynamical charge, $\hat{Q}_j = \hat{a}_j^{\dagger} \hat{a}_j - \hat{b}_j^{\dagger} \hat{b}_j$, is basically the difference between the particle-antiparticle number. 
In the absence of any background static charges, the physical sector is spanned by the states satisfying $\hat{G}_j =0$ for all values of $j$.  
Using this constraint imposed by the Gauss law, we can integrate-out the gauge-fields for a chain  with open-boundary condition using the following transformation,
\begin{equation}
 \left[\prod_{l < j} \hat{U}_l \right] \hat{a}_j \rightarrow \hat{a}_j, \ \left[\prod_{l < j} \hat{U}^{\dagger}_l \right] \hat{b}_j \rightarrow  \hat{b}_j,
\hat{L}_j = \sum_{l \leq j} \hat{Q}_l,
\label{eq:integrateout}
\end{equation}
\normalsize 
where the background static field at the left of the chain has been considered to be zero.

\begin{figure}
\includegraphics[width=0.7\linewidth]{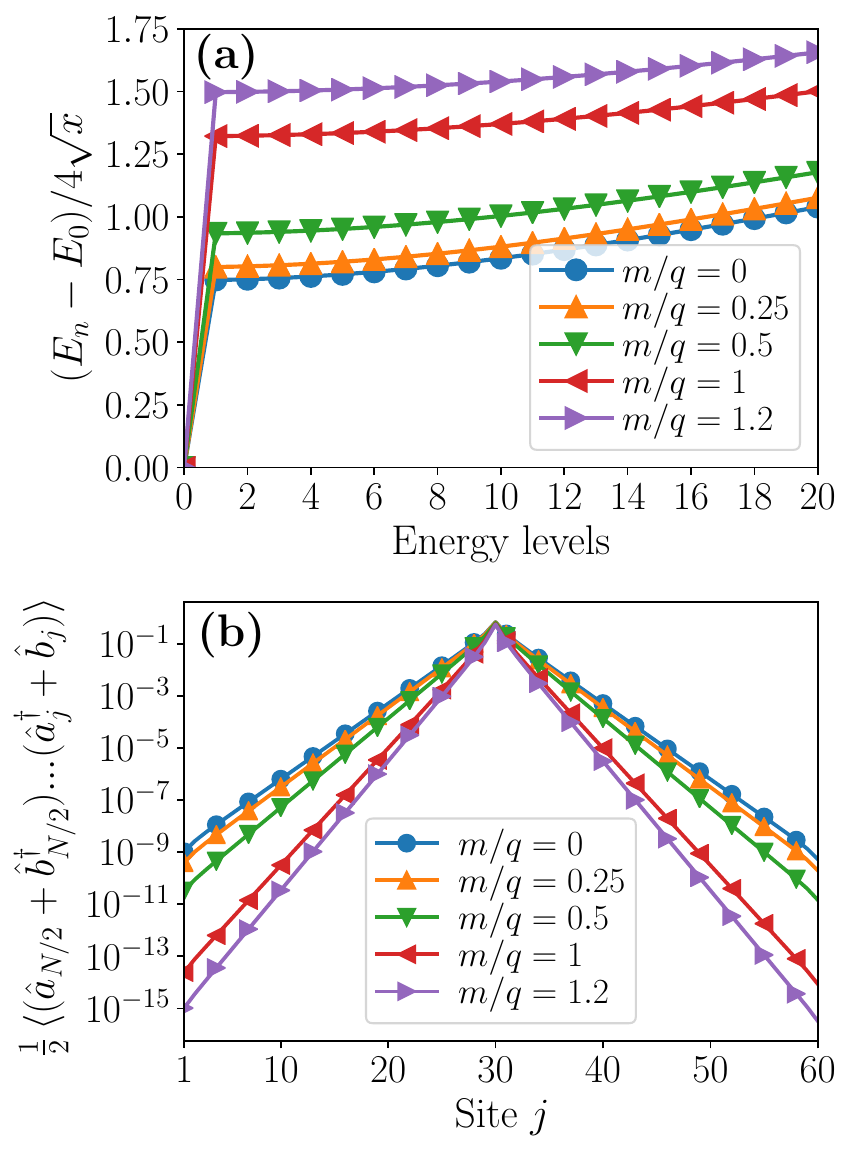}
\caption{{(Color online.) Energy levels of 20 low-lying states for different values of $m/q$ as obtained from the DMRG calculation. 
Clearly, the ground states are well-separated from the energy bands even in the massless case due the matter-gauge interaction, responsible for opening of mass-gaps.
(b) Exponentially decaying correlation in the ground states of the system. Here, we plot the long-range correlation, $\frac{1}{2}\braket{(\hat{a}_{N/2} + \hat{b}^{\dagger}_{N/2})...(\hat{a}^{\dagger}_{j} + \hat{b}_{j})}$, as a function of distance $j$ from the middle of the chain, i.e., site $N/2$. Note that the ellipsis $...$ is to be replaced by proper Wilson lines to make the long-range operators gauge-invariant}.
}
\label{fig:gr}
\end{figure}

\begin{figure}
\includegraphics[width=0.8\linewidth]{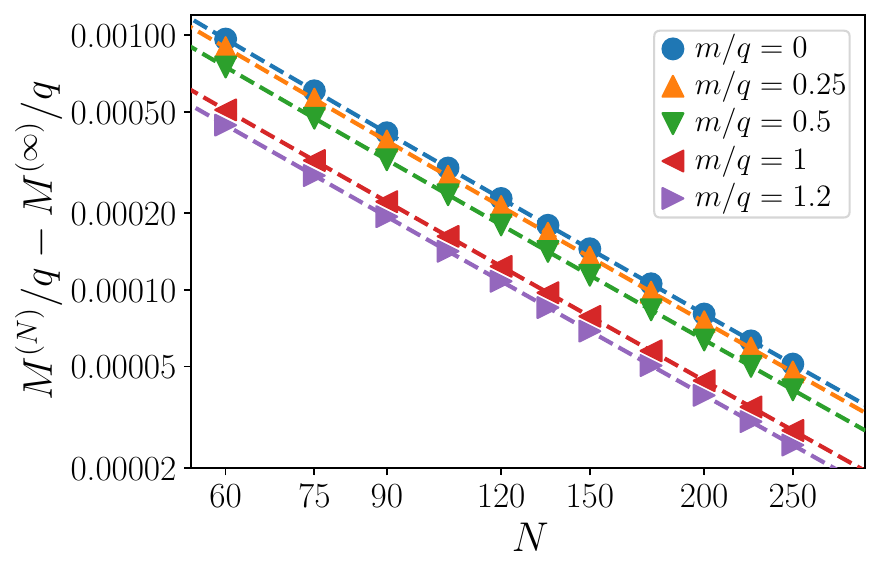}
\caption{({Color online.) Finite-size scaling of mass-gap for different values of $m/q$. Here, we fit the values of mass-gaps obtained for different system-sizes to Eq. \eqref{eq:scaling} to obtain the thermodynamic limits. The symbols represent the numerical data,  the dashed lines show the fitted results. Error of these fits are of the order of $10^{-7}$ (for smaller masses) to $10^{-9}$ (for higher masses).
All quantities plotted are dimensionless.}}
\label{fig:scaling}
\end{figure}

\subsection{Ground-state, mass-gap, and energy spectrum}

\subsubsection{Energy spectrum in the non-interacting scenario}
{If we drop the $\sum_j \hat{L}_j^2$ term from the Hamiltonian (Eq. \eqref{eq:Hamil4}), then the system is basically the discretized version of free (complex) Klein-Gordon (KG) theory in $1+1$ dimension, which can be solved exactly. After successive application of Fourier and Bogoliubov transformations, the non-interacting KG Hamiltonian,  corresponding to the Hamiltonian of Eq.~\eqref{eq:Hamil4}, boils down to
\begin{eqnarray}
\hat{H}_{free} = \sum_k \omega_k \left( \hat{a}_k^{\dagger} \hat{a}_k + \hat{b}_k^{\dagger} \hat{b}_k + 1 \right),
\end{eqnarray}
where the doubly degenerate spectrum of the KG system is given by the 
dispersion relation
\begin{eqnarray}
\omega_k = 2 \sqrt{x m^2 / q^2+ 2 x^2 (1 - \cos k a)},
\end{eqnarray}
for both the particles and the antiparticles.
This dispersion relation is transformed into the relativistic one in the continuum limit as $\lim_{a \rightarrow 0} \frac{a q^2}{2} \omega_k = \sqrt{k^2 + m^2}$. 
Note that for very low momentum excitations, i.e., $k a \ll 1$,  the system also possesses Lorentz invariant dispersion even for finite size-size with finite lattice-spacing. 
Clearly, the ground state of the non-interacting system, which is the bare vacuum of the KG theory, is gapless for massless bosons and gapped otherwise,  and the energy gap is given by, $\Delta E = 2 \times 2 \sqrt{x} m/q$, where a factor of $2$ is needed as the excitations are in the from of `free' boson-antiboson  pairs.}

\subsubsection{Emergence of mass-gap and binding energy}

{In presence of $\sum_j \hat{L}_j^2$ term in the Hamiltonian, 
excitations come in the form of bound particle-antiparticle pairs (mesons), and 
a finite mass-gap is generated  due to the matter-gauge coupling.
The mass-gap, i.e., the difference in energies between the ground and first excited states $M/q = (E_1-E_0)/4\sqrt{x}$,  is always larger than $m/q$, where the extra energy, $E_B/q = M/q - m/q$ arises in the from of binding energy required to tether particle-antiparticle pairs into mesons. 
In Fig.~\ref{fig:gr}(a), we plot energies (as measured from the ground state energy)
of 20 low-lying states as obtained from the DMRG calculation that shows the existence of mass-gap in the system.
Due to this mass-gap, the ground state of the system is finitely correlated.
Fig.~\ref{fig:gr}(b) shows that long-range correlations decay exponentially with the distance in the ground state.}

{We move on with our analysis by extracting the information about mass-gap $M/q$ and the corresponding binding-energy, $E_B/q = M/q - m/q$, in the thermodynamic limit by finite-size scaling. 
Due to the open boundary condition, the mass-gaps will receive a kinetic energy correction with a leading contribution being $O(1/N^2)$  as we approach the thermodynamic limit  \cite{hamer_prd_97}. Here, we consider the contributions upto $O(1/N^3)$ perturbation term, so that the scaling is given by
\begin{eqnarray}
M^{(N)}/q = M^{(\infty)}/q + \alpha_1/ N^{2} + \alpha_2/N^3,
\label{eq:scaling}
\end{eqnarray}
where $\alpha_1$ and $\alpha_2$ are dimensionless constants.
We obtain the values of the mass-gap from DMRG calculation for different system-sizes $N \in [60, 250]$, and numerically fit the data to Eq. \eqref{eq:scaling} (see Fig. \ref{fig:scaling}) to get the mass-gap in the thermodynamic limit. In Table \ref{tab:binding}, we list the values of mass-gaps and corresponding binding energies in the thermodynamic limit for different values of $m/q$.}

\begin{table}
\begin{tabular}{|c|c|c|}
\hline
$m/q$ & Mass-gap, $M/q$ & Binding energy, $E_B/q$ \\
\hline \hline
0 & 0.74688 & 0.74688 \\

0.25 & 0.79872 & 0.55872\\

0.5 & 0.93303 & 0.53303 \\

1 & 1.32112 & 0.32112 \\

1.2 & 1.49636 & 0.29636 \\
\hline
\end{tabular}
\caption{{Mass-gaps and binding energies for $x=2$ in the thermodynamic limit as extracted from the finite-size scaling.}}
\label{tab:binding}
\end{table}

\begin{figure}
\includegraphics[width=\linewidth]{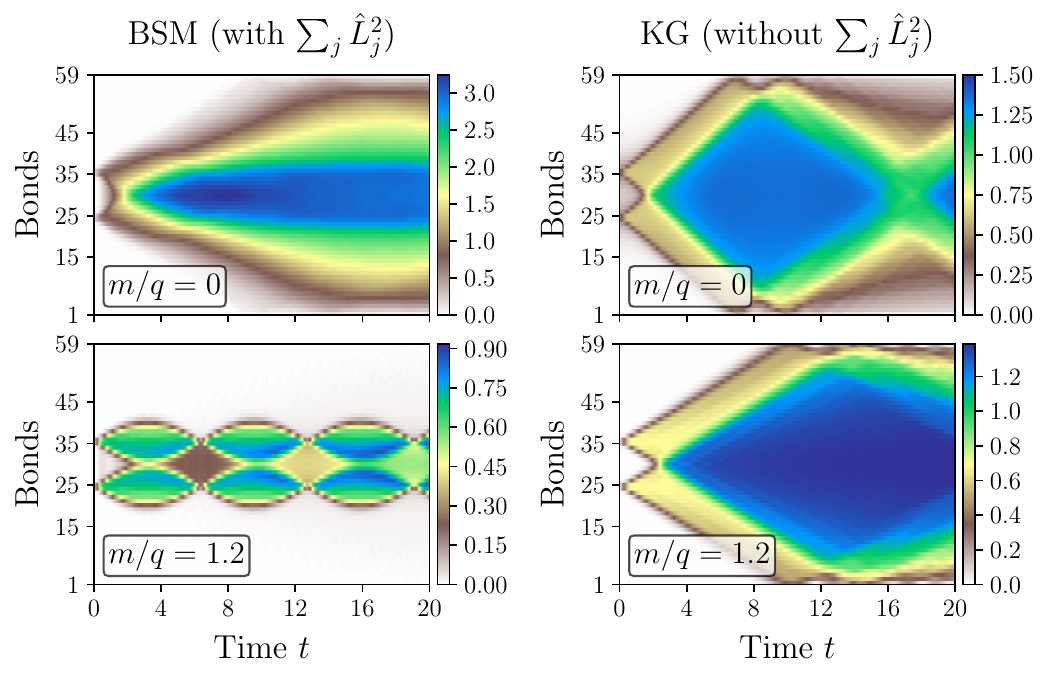}
\caption{(Color online.) Dynamics of entanglement in the BSM (left column) and in the non-interacting KG system (right column). We plot the entanglement entropy of the time-evolved state after subtracting the entropy of respective ground state $\ket{\Omega}$ and consider $R=5$.  
}
\label{fig:dyn2}
\end{figure}

\subsection{Entanglement dynamics} 
{In the main text, we have presented the time profile for classical and distillable part of the entanglement entropy. Here, we supplement those results with the profile of the total entanglement entropy $\mathcal{S}_j(t)$ measured across every bond $j$ and compare it with what is observed in a  non-interacting KG system, obtained by dropping $\sum_j \hat{L}^2_j$ term from the Hamiltonian.} 

{Fig.~\ref{fig:dyn2} shows $S_j(t)$ for the interacting BSM (left column), and the  the non-interacting KG model (right column) for 
all bonds $j$.  
For the KG model the  entanglement spreads linearly with a light-cone structure as predicted by the pseudo-particle picture. For a given $j\ll N/2$, it initially increases ballistically with time and saturates to a value proportional to the volume of the region as predicted by \cite{calabrese_jstatmech_2005} and is recently discussed in the context of generalized thermalization \cite{rigol_nature_2008, vidmar_jstatmech_2016}. The particles bouncing off the boundaries induce the observed recurrences.}   

{In the BSM, the spread of the entanglement is strongly modified by the effects of confinement (left column). Initially its spreading slows-down, and only starts to spread ballistically in correspondence to the radiation of free mesons for lighter masses. Furthermore, most of the entanglement is contained in the region that is initially occupied by the confined bosonic matter, and persists there even long after the concentration of bosons in the bulk disappears at around $t\simeq 10$. 
On the other hand, the concentration of entanglement never leaks into the deconfined domain for heavier bosons, e.g., $m/q = 1.2$. Such unusual dynamics of entanglement gives us yet another indicator of strong reluctance 
towards thermalization.}

\begin{figure}
\includegraphics[width=\linewidth]{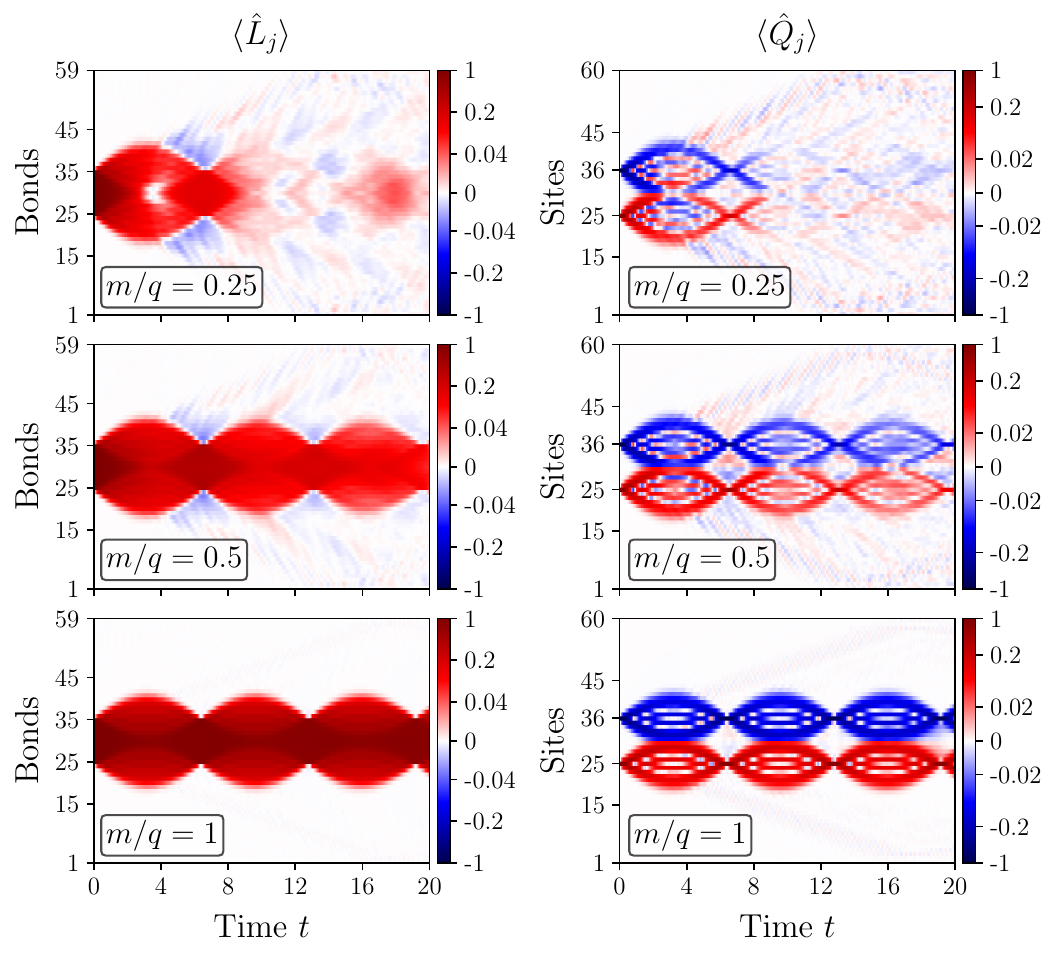}
\caption{(Color online.) Dynamics of electric field $\braket{\hat{L_j}}$ (left column) and dynamical charge $\braket{\hat{Q}_j}$ (right column)
for $m/q = 2.5$ (top row), $0.5$ (middle row), and $1$ (bottom row) and $R=5$. 
}
\label{fig:dyn1_supple}
\end{figure}

\begin{figure}
\includegraphics[width=\linewidth]{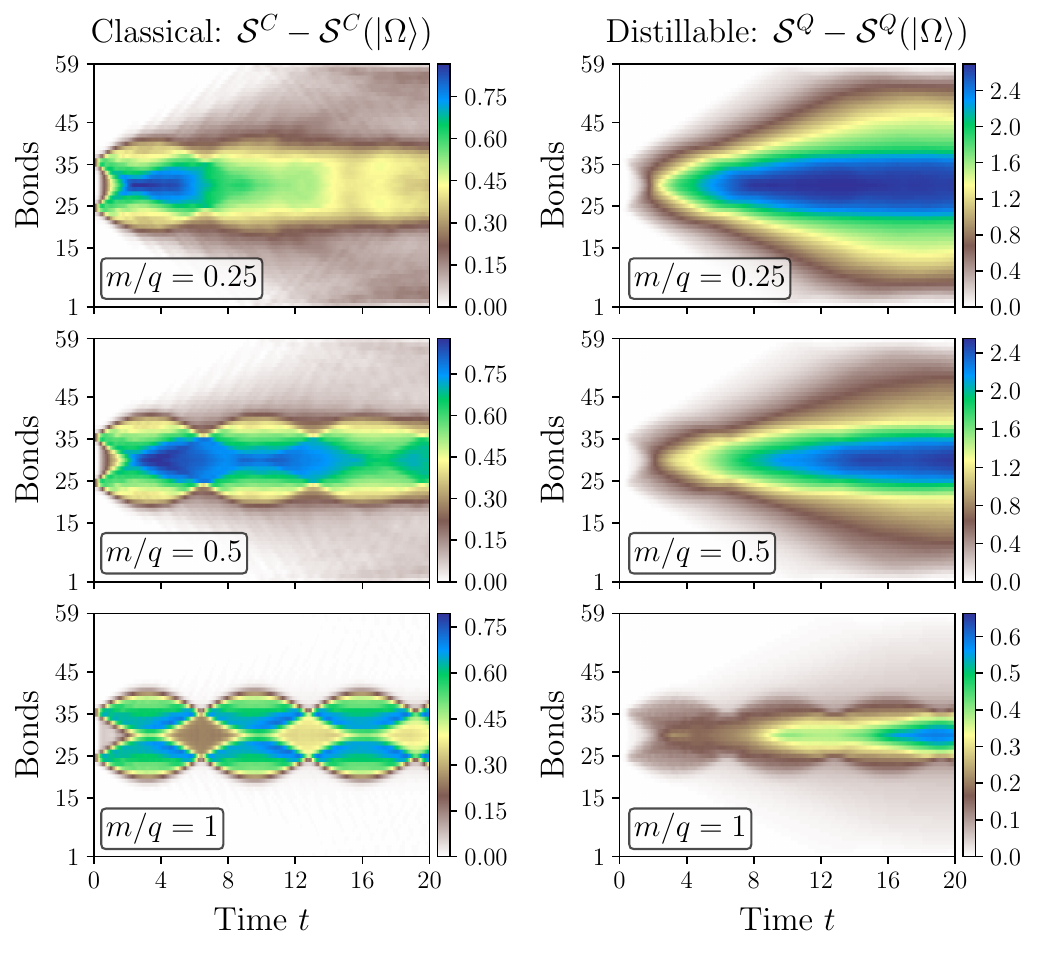}
\caption{(Color online.) Time-evolution the classical part $\mathcal{S}^C$ of entanglement entropy (left column) and the distillable entanglement entropy $\mathcal{S}^Q$ (right column)
for $m/q = 2.5$ (top row), $0.5$ (middle row), and $1$ (bottom row) and $R=5$. 
}
\label{fig:dyn2_supple}
\end{figure}

\subsection{Dynamics for intermediate boson masses}
{In the main text, we have presented the out-of-equilibrium dynamics for two extreme values of boson mass, namely $m/q = 0$ and $1.2$. Here, we supplement those results by showing the dynamics for intermediate values of $m/q$.}
Fig. \ref{fig:dyn1_supple} shows the dynamics of both the gauge sector ($\hat{L}_j$) and the charge sector ($\hat{Q}_j$) for boson masses $m/q = 0.25$, $0.5$, and $1$ for $R=5$. Clearly, as the mass increases the confining behavior of the dynamics becomes stronger, as the coherent oscillation of the confined core lasts longer. For example, there is no string-inversion for $m/q = 0.25$ before $t = 8$, and  the string does not break in the bulk for $m/q = 0.5$. On the other hand, we already reach the heavy boson limit with $m/q = 1$.

The classical ($\mathcal{S}^C$) and the distillable ($\mathcal{S}^Q$) parts of entanglement entropy shows similar features for these intermediate masses (Fig. \ref{fig:dyn2_supple}). The distinction between the deconfined domain and the confined core, as perceived from the classical part,  becomes much more pronounced for $m/q = 0.25$ and $0.5$ than the massless scenario depicted in the main text.

\subsection{Area-law of entanglement entropy in the confined domain}
\begin{figure}
\includegraphics[width=\linewidth]{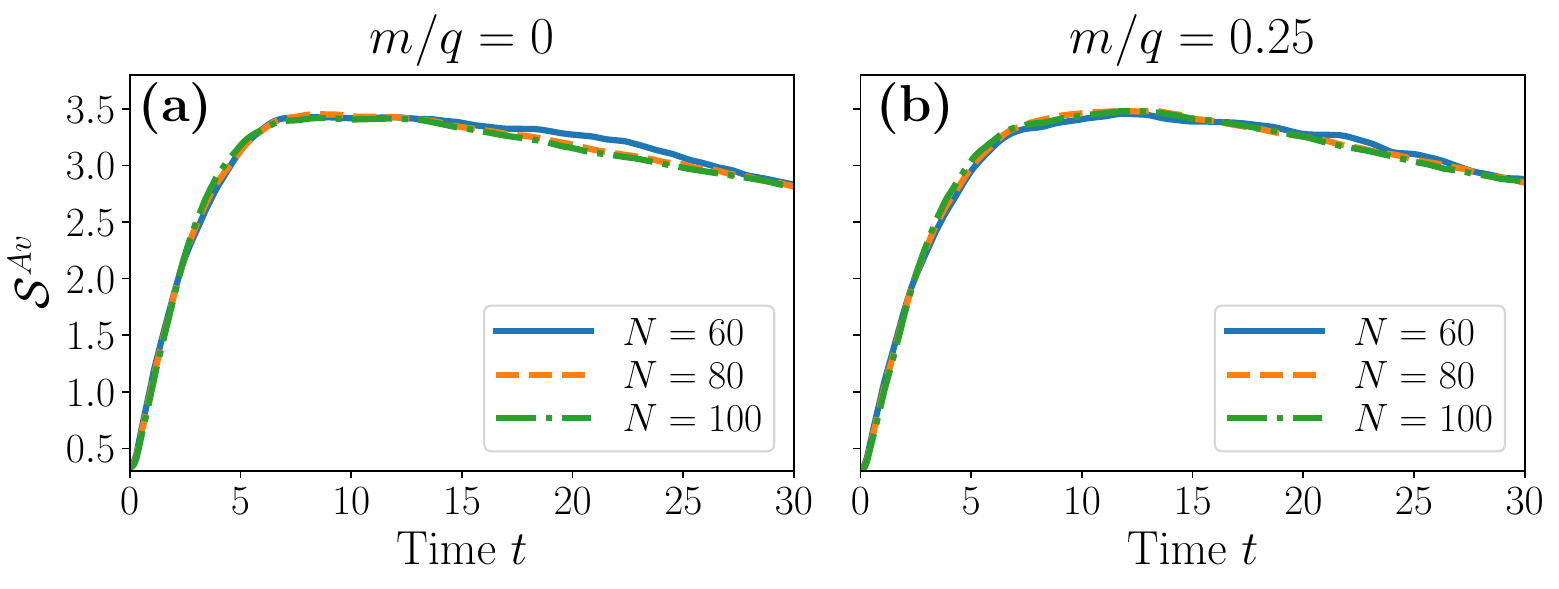}
\caption{(Color online.) Evolution of average entropy in the confined domain for different system sizes. Here we consider $R = N/10$ so that it grows proportional to the system size.
}
\label{fig:entAv}
\end{figure}

As mentioned in the main text, we follow the entanglement dynamics for system sizes $N = 60$, $80$, and $100$ with $R=N/10$ so that the initial string length increases with the system size. The average entropy in the confined domain, i.e., 
\begin{eqnarray}
\mathcal{S}^{Av} = \frac{1}{2 R + 1} \sum_{j = N/2 - R}^{N/2 + R} \mathcal{S}_j
\end{eqnarray}
follows the area-law of entanglement throughout the dynamics as it remains (almost) invariant with the system size as shown in Fig. \ref{fig:entAv} for $M/q = 0$ and $0.25$. Here, the entropy grows rapidly and reaches its maximum value at $t \approx 10$ and then starts to decay slowly similar to the situation of larger masses at much later times that we have seen in the main text.

\begin{figure}
\includegraphics[width=0.6\linewidth]{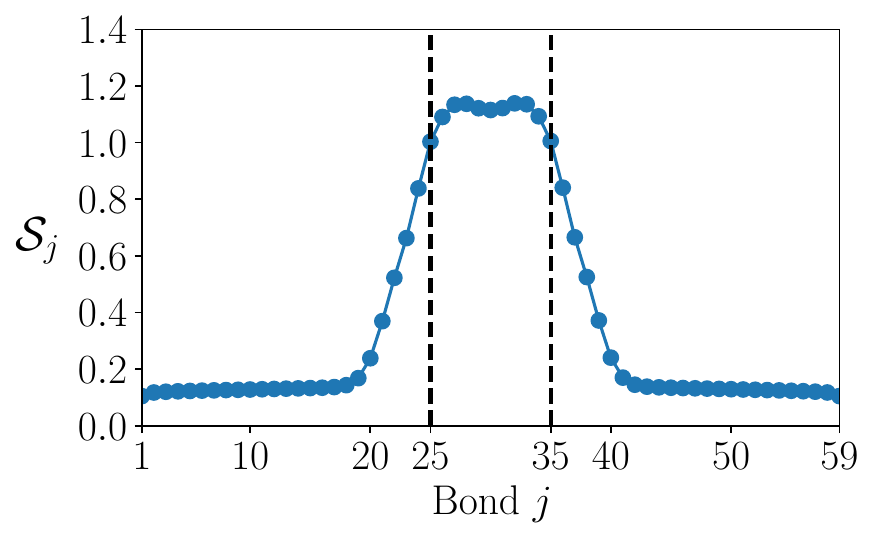}
\caption{(Color online.) The entanglement profile for $m/q=1.2$ at time $t=200$. Here, we consider $N=60$ and $R=5$. The region inside the dashed lines refers to the central confined region. 
}
\label{fig:entprof}
\end{figure}

The area-law of entanglement entropy can be also perceived from the entanglement profile of the systems at long times. In Fig. \ref{fig:entprof} we depict the entanglement profile for $m/q = 1.2$ at $t=200$ for $N=60$ and $R=5$. Clearly, the entropy in the central confined region remains almost invariant with the size of the bipartition, and the region just outside this central region starts to show a linear increase of entanglement entropy with the size of the bipartition.

\begin{figure*}
\includegraphics[width=0.7\linewidth]{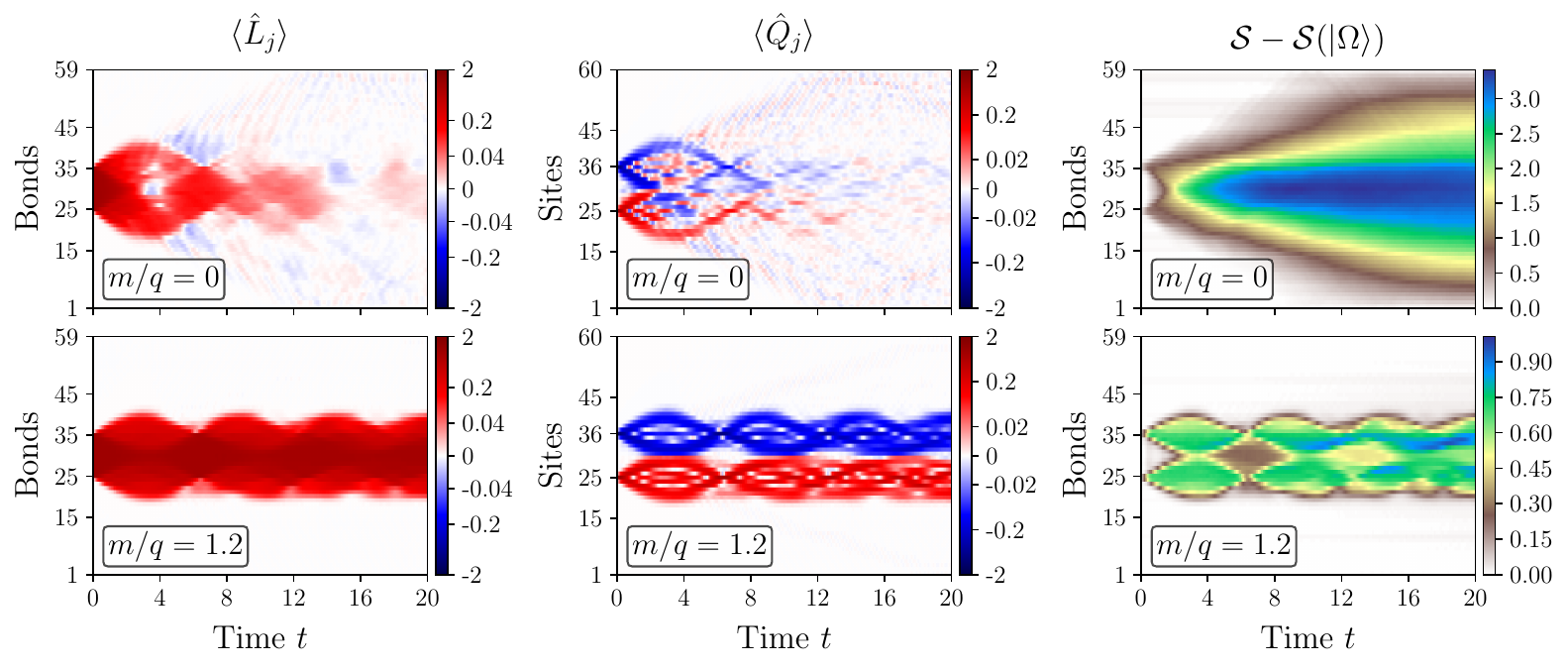}
\caption{{(Color online.) Dynamics of the electric field $\braket{\hat{L}_j}$, the dynamical charge $\braket{\hat{Q}_j}$, and entanglement entropy $\mathcal{S}_j$ for $R = 5$ and $m/q=0$ under random on-site disorder. Random on-site potential $\sum_j \mu_j \left(\hat{a}_j^{\dagger} \hat{a} + \hat{b}_j^{\dagger} \hat{b} \right)$ has been added to the system where $\mu_j$ has been chosen randomly in between $[-1/2, 1/2]$.}}
\label{fig:random}
\end{figure*}

\subsection{Robustness of confining dynamics under random noise}
{Here, we show that the confining dynamics described in the main text is robust under random noises that may be present in the system. For that purpose, we add on-site random potential $\mu_j \sum_j \left(\hat{a}_j^{\dagger}\hat{a}_j +  \hat{b}_j^{\dagger}\hat{b}_j\right)$ to the Hamiltonian, where $\mu_j$ is chosen randomly from $\left[-1/2, 1/2\right]$. 
Fig. \ref{fig:random} shows an instance for one such dynamical behavior for the massless case. Although the profile of the light cones get deformed due to random disorder, 
the confining dynamics can be easily grasp from the bending of the trajectories of the bosons. More importantly, the entanglement entropy remains concentrated on the central region like in the clean case, thereby indicating a strong memory effect and thus lack of thermalization.}

\begin{figure*}
\includegraphics[width=0.7\linewidth]{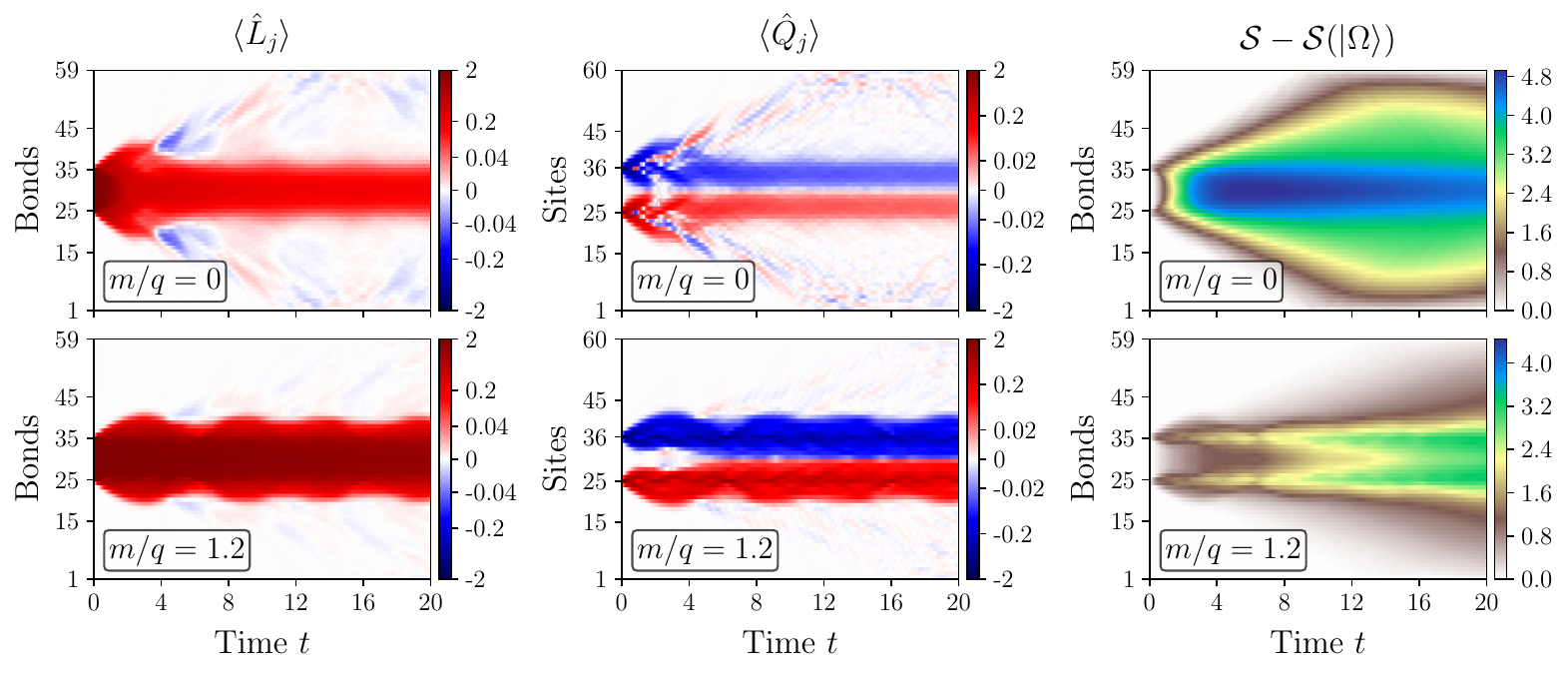}
\caption{{Dynamics of the electric field $\braket{\hat{L}_j}$, the dynamical charge $\braket{\hat{Q}_j}$, and entanglement entropy $\mathcal{S}_j$ for the initial state $\ket{\psi'(t=0)} = \mathcal{N}'\hat{M}^2_R\ket{\Omega}$, with $\mathcal{N}'$ being the normalization constant and $R = 5$.}}
\label{fig:LQS2}
\end{figure*}

\subsection{Confining dynamics at higher energies}
{We also probe the dynamics at a larger energy than the scenario presented in the main text. For that, we excite the ground state by acting the non-local string operator $\hat{M}_R$ 
twice, such that the initial state becomes, $\ket{\psi'(t = 0)} = \mathcal{N}' \hat{M}_R^2 \ket{\Omega}$, with $\mathcal{N}'$ being the
normalization constant. Semiclassically, this initial state has twice the extra energy than the previous scenario. In Fig. \ref{fig:LQS2}, we depict such dynamics of the electric field $\braket{\hat{L}_j}$, the dynamical charge $\braket{\hat{Q}_j}$, and entanglement entropy $\mathcal{S}_j$ for $m/q = 0$ and $1.2$.
In spite of having more energy than the previous case, the memory effect becomes more prominent, as the concentration of bosons remains localized in the central region for much longer time.}

\begin{figure}[h]
\includegraphics[width=0.7\linewidth]{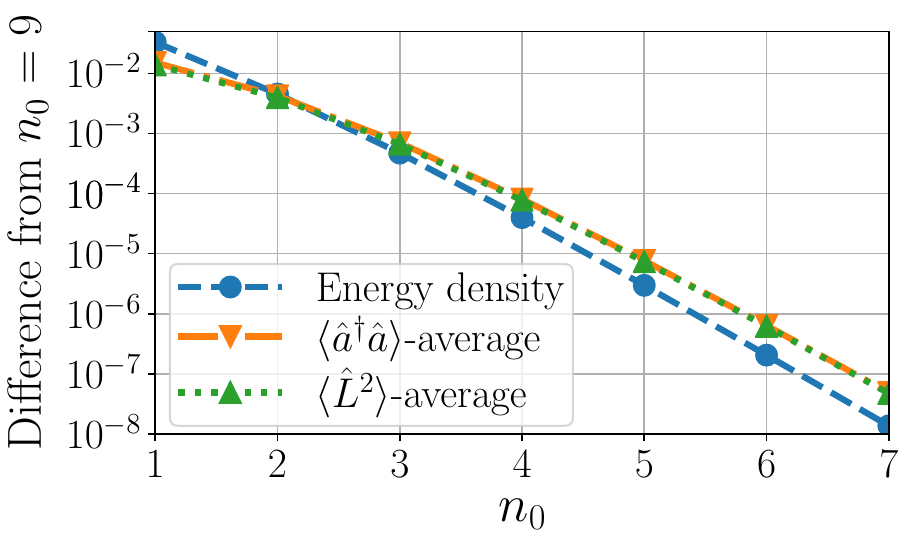}
\caption{(Color online.) Dependence of physical quantities on the boson number cutoff $n_0$ in the ground state of the system for $m/q= 0$. Here we plot the difference between the values of different observables computed for different values of $n_0$ from $n_0=9$. Clearly, for $n_0=5$ the error due to the truncated bosonic Hilbert space falls below $10^{-5}$.}
\label{fig:nconv}
\end{figure}

\begin{figure*}[t]
\includegraphics[width=0.9\linewidth]{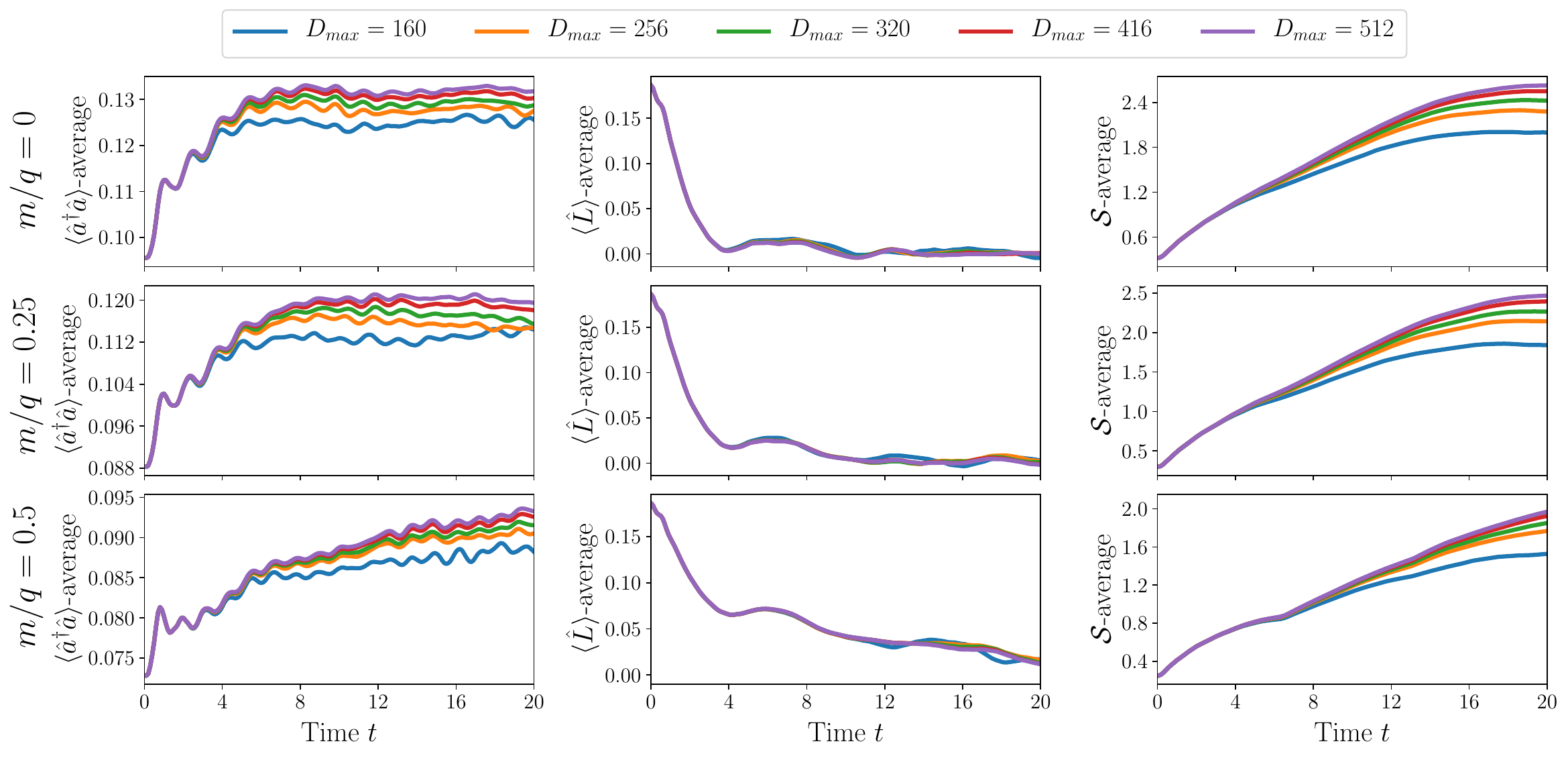}
\caption{(Color online.) Time-evolution of different physical observables using hybrid TDVP for different bond-dimensions. We plot the averages of $\braket{\hat{a}^{\dagger} \hat{a}}$ (left column), $\braket{\hat{L}}$ (middle column), and the entanglement entropy $\mathcal{S}$ (right column) as function of time upto $t=20$ for $m/q = 0$, $0.25$, and $0.5$.}
\label{fig:tconv}
\end{figure*}

\subsection{Details about tensor network simulation}
We use matrix product states (MPS) \cite{schollwock_aop_2011, orus_aop_2014} ansatz with open boundary condition to simulate states of the system, where we integrate-out the gauge fields using the Gauss law. Due to tracing-out of the gauge fields, we do not need to use gauge-invariant tensor network \cite{tagliacozzo_prx_2014, buyens_prl_2014, silvi_njp_2014, kull_aop_2017} for our calculations. However, we use global $U(1)$ symmetry \cite{singh_pra_2010, singh_prb_2011} corresponding to the conservation of the total dynamical charge, $\sum_j \hat{Q}_j$, and obviously we work only in the $\sum_j \braket{\hat{Q}_j} = 0$ sector.
The maximum number bosons ($n_0$) per site for each species has been truncated to 5, resulting in a physical dimension of 36 on each site. This truncation is justified as the densities of the bosons never cross $\sim 1.5$ throughout our simulation. We confirm this by checking the convergence of several observables with respect to $n_0$ in the ground-state of the model
in Fig. \ref{fig:nconv}, where we show that for $m/q = 0$ the errors due to this truncation is below $10^{-5}$ for $n_0 = 5$. 
One important thing to mention here is that it is also possible to separate-out two types of bosons to odd and even lattice-sites respectively maintaining global $U(1)$ symmetry, such that physical dimension on each site only grows linearly with $n_0$. This will definitely increase efficiency of the simulation for a given MPS bond dimension. However, as two types of bosons sitting on a same site are \emph{strongly} correlated, such a method of separating them out using truncated bonds will incur much more errors, especially in the time-evolution, and needs much larger bond dimension to get converged results. 

To find the ground state of the system, first we use two-site density matrix renormalization group (DMRG) \cite{white_prl_1992, white_prb_1993, schollwock_rmp_2005} upto a maximum bond dimension $D_{max} \leq 100$, so that largest SVD truncation error with $D_{max}$ remains below $10^{-12}$.  
After that we switch to one-site variational optimization (``one-site DMRG") \cite{white_prb_2005, schollwock_aop_2011} for more stringent convergence within the MPS manifold given by $D_{max}$.

To obtain low-lying excited states, we employ the same method, where we shift the Hamiltonian each-time by a suitable weight factor multiplied with the projector of the previously found state, i.e., to find the $n^{th}$ excited state $\ket{\psi_n}$, we search for the ground state of the shifted Hamiltonian, 
\begin{equation}
\hat{H}' = \hat{H} + W \sum_{m=0}^{n-1} \ket{\psi_m}\bra{\psi_m},
\end{equation}
 where $W$ should be guessed to be sufficiently larger than $E_n - E_0$.
In this scenario, $D_{max} = 100$ is not always sufficient to reduce the SVD truncation error below $10^{-12}$ and that is why  convergence from the one-site variational optimization procedure becomes absolutely necessary.

Time-evolution using MPS ansatz (tensor network in general) is always tricky, error-prone, and therefore must be dealt with caution, as entanglement entropy grows ballistically in the dynamics, which, in turn, demands larger and larger MPS bond dimension.  Recently, to tackle such issues,  the time-dependent variational principle (TDVP) algorithm  \cite{haegeman_prl_2011, koffel_prl_2012, haegeman_prb_2016} has been developed, which has been argued to be much less error-prone than earlier methods, e.g., time evolving block decimation (TEBD), within a given bond dimension \cite{paeckel_arxiv_2019}.
Here we employ ``hybrid'' TDVP with step-size $\delta t = 0.01$, where we first use two-site version of TDVP to dynamically grow the bond dimension upto $D_{max} = 512$. When the bond dimension in the bulk of the MPS is saturated to $D_{max} = 512$, we switch to the one-site version to avoid any error due to SVD truncation. This hybrid method of time-evolution using TDVP has been argued to incur much less error than other known methods \cite{paeckel_arxiv_2019, goto_prb_2019}. It is noteworthy to mention here that since we use properly converged Lanczos exponentiation \cite{hochbruck_siam_1997} in TDVP simulations, different step-sizes do not alter the results.
To be assured of the trustworthiness of our simulations, we also perform TDVP simulations for $D_{max}= 160$,  $256$, $320$, and $416$ and check the convergence of different observables with respect to different $D_{max}$ (see Fig. \ref{fig:tconv} for the case of $N=60$ and $R=5$). Clearly, upto $t \approx 4$, all the graphs, including $D_{max} = 160$ simulations, are converged, even when tallied in the light of entanglement entropy $\mathcal{S}$, which is believed to behave much worse in truncated bond dimensions. On the other hand, $D_{max} = 416$ data remain satisfactorily close to $D_{max} = 512$ curves throughout the time window, showing the reliability of our simulation with the bond dimension $D_{max}= 512$.  For heavier masses, e.g.,  $m/q = 1.2$ (not shown in the figure), all the quantities are converged for every bond dimension considered here.

\bibliography{supple_bsm_jz.bbl}